\documentclass[reprint,superscriptaddress,amsmath,amssymb,aps,floatfix,pra]{revtex4-1}
 \pdfoutput=1

\usepackage[normalem]{ulem}
\usepackage{graphicx}
\usepackage[dvipsnames]{xcolor}
\usepackage{verbatim}
\usepackage{mje,dsfont}
\usepackage{hyperref}
\usepackage[utf8]{inputenc}
\usepackage{upgreek}
\newcommand{\IsoWigner}{equiumbral\xspace}
\newcommand{\QSERG}{Quantum Systems Engineering Research Group, Department of Physics, Loughborough University, Loughborough, Leicestershire LE11 3TU, United Kingdom}
\newcommand{\Wolfson}{The Wolfson School, Loughborough University, Loughborough, Leicestershire LE11 3TU, United Kingdom}

\newcommand{\br}[1]{\left({#1}\right)}

\newcommand{\Tab}[1]{Table~(\ref{#1})}

\definecolor{PatGreen}{rgb}{0.0, 0.4, 0.1}

\newcommand{\noniso}{non-isomorphic~}

\def\Bid{\mathds{1}}

\begin{document}

\title{On quantum invariants and the graph isomorphism problem}

\author{P.~W.~Mills}
\affiliation{\QSERG}
\author{R.~P.~Rundle}
\affiliation{\QSERG}
\affiliation{\Wolfson}
\author{J.~H.~Samson}
\affiliation{\QSERG}
\author{Simon J. Devitt}
\affiliation{Centre for Quantum Software \& Information (QSI), Faculty of Engineering \& Information Technology, University of Technology Sydney, Sydney, NSW, 2007, Australia.}
\affiliation{Turing inc., Berkeley, CA, 94701 USA}
\author{Todd Tilma}
\affiliation{Tokyo Institute of Technology, 2-12-1 Oookayama, Meguro-ku, Tokyo 152-8550, Japan}
\affiliation{\QSERG}
\author{V.~M.~Dwyer}
\email{v.m.dwyer@lboro.ac.uk}
\affiliation{\QSERG}
\affiliation{\Wolfson}
\author{Mark~J.~Everitt}
\email{m.j.everitt@physics.org}
\affiliation{\QSERG}

\date{\today}

\begin{abstract}
Three new graph invariants are introduced which may be measured from a quantum graph state and form examples of a framework under which other graph invariants can be constructed. Each invariant is based on distinguishing a different number of qubits. This is done by applying alternate measurements to the qubits to be distinguished.
The performance of these invariants is evaluated and compared to classical invariants. 
We verify that the invariants can distinguish all non-isomorphic graphs with 9 or fewer nodes.
The invariants have also been applied to `classically hard' strongly regular graphs, successfully distinguishing all strongly regular graphs of up to 29 nodes, and preliminarily to weighted graphs. 
We have found that although it is possible to prepare states with a polynomial number of operations, the average number of preparations required to distinguish \noniso graph states scales exponentially with the number of nodes. 
We have so far been unable to find operators which reliably compare graphs and reduce the required number of preparations to feasible levels.

\end{abstract}
\maketitle
\section{Introduction}

A graph is  a set of  nodes connected by edges, and two graphs are termed isomorphic if one may be obtained from the other by permuting the labels of their nodes~\cite{Harary1972}. The question of whether two graphs are isomorphic is the so-called graph isomorphism (GI) problem~\cite{Gould1988}: a computationally hard problem that is not just of academic interest, but is also central to a number of areas critical to industry. Some of the more obvious of these include the following: the integrated circuit industry requires designs to pass a key (layout versus schematic) verification check, which compares the transistor network delivered by the Logic Synthesizer to that extracted from the Place and Route engine~\cite{Roberts2017}. In the field of image recognition, including registration problems in computer vision~\cite{berg2005} and medical imaging (for example automated histology analysis~\cite{McCann2015}), graphs are used as effective structural descriptors due to their ability to represent relational information in which nodes are associated to image components and edges to the relationships between them.  In the field of cybersecurity, the control flow graphs of a number of worms have been analysed as a detection technique~\cite{Kruegel2005}. Perhaps less obvious applications of GI involve financial fraud detection, banking risk management, legal precedents, fault detection, and even zero-knowledge proofs~\cite{QCAPS,Blum1986}.
Some industry experts have estimated that the global worth of a number of these sectors will grow up to \$12Bn within the next ten years,
making progress in solving the GI problem an important industrial as well as academic challenge~\cite{QCAPSRep}. 

Complexity arises here as, even restricted to simple undirected graphs without loops, the number of non-isomorphic graphs increases at least exponentially with the number of nodes~\cite{Harary1972}. The combination of a number of different methods has resulted in classical algorithms that are efficient for many graphs~\cite{Ullmann1976,McKay201494,VF2}. However, there still exist a large number of important graphs for which solutions do not currently exist~\cite{QCAPS,Neuen2017}. As a result, new contributions to graph isomorphism, such as those that might be offered by quantum computing, would add real value, even if they only deal with these difficult cases. 

While the problem is computationally hard, simple methods such as edge counting or spectral comparisons can resolve many cases efficiently. These comparisons rely on the fact that such properties (known as graph invariants~\cite{lovasz2012large}) are shared by all isomorphic graphs. The current best classical algorithm for the general case is due to  Weisfeiler and Lehman~\cite{Weisfeiler1968,W1968Trans,Weisfeiler1976} implemented in the Nauty algorithm~\cite{McKay201494} which is able to solve many cases of practical interest in polynomial time. However, a general solution with polynomial scaling does not exist, and recent unpublished work indicates that the problem is classically solvable in quasi-polynomial time, implying that the GI problem is between polynomial and exponential in its complexity~\cite{1512.03547,1710.04574}. 
It is worth noting here that Nauty performs poorly for strongly regular graphs, graphs which the quantum algorithms developed here are able to distinguish (up to the 29 node graphs tested). This demonstrates that our algorithms are fundamentally different from (at least two dimensional) Weisfeiler-Lehman methods ~\cite{Cai1992,Douglas2011}. 

Two factors support the existence of an efficient algorithmic quantum solution for the graph isomorphism problem. The first is that the complexity of the GI problem is similar to that of integer factorization~\cite{CComp2009}, a problem which can be solved in polynomial time on a quantum computer using the Shor algorithm~\cite{Shor1997}. The second reason follows from the existence of an adiabatic quantum-annealing method by Gaitan and Clark~\cite{Gaitan2014}, which already solves the GI problem. As the algorithm is adiabatic, the method's complexity is unknown. However since adiabatic and algorithmic quantum computing are known to be equivalent~\cite{Aharonov2004}, the adiabatic method guarantees the existence of a quantum solution. 
We also note here the work of Wang \textit{et al.\ }~\cite{Douglas2008,Berry2011} applying quantum walks to the GI problem; these methods have successfully distinguished classes of strongly regular graphs with up to 64 nodes. However, their method (like many novel approaches~\cite{Smith2010}) has been shown to be equivalent to the Weisfeiler-Lehman algorithm~\cite{Douglas2011}. 

With many methods equivalent to Weisfeiler-Lehman's (W-L), effective general invariants unrelated to W-L's are of significant interest to the field. Any W-L based method cannot distinguish Cai-F\"urer-Immerman graphs~\cite{Cai1992,Furer1987}. The ability to distinguish such graphs would therefore demonstrate our algorithms are distinct from all dimensions of the W-L method.
Whilst we have been unable to analyse such large graphs due to the scaling of our classical simulations, we believe our method works in a fundamentally different manner since it is not iterative and exploits the exponential resources available on a quantum computer to measure group elements, which are guaranteed to be different for non-isomorphic graphs.

Motivated by these factors we introduce a number of graph invariants which were designed to exploit the exponential resources of a quantum computer. These and the classical invariants are compared below for their efficacy in distinguishing graphs, as a function of graph node number. 
We have found that our invariants are better at distinguishing graphs than several classical methods in the sense that a higher proportion of graphs give a unique result. 
Indeed, in the worst case, the proportion of graphs which the quantum invariants cannot distinguish appears to tend to zero. In the particular case of strongly regular graphs, the quantum invariants are able to distinguish all graphs with fewer than 30 nodes.
Furthermore, we find that two of our quantum measures allow all the non-isomorphic graphs that we have been able to consider in this work to be distinguished.

This work presents a family of new graph invariants which could potentially be extended to form part of a practical solution. However, it also highlights the difficulty of obtaining information from a large quantum system, even in the ideal case.


\section{Constructing Graph States}
\label{sec:Constructing}

To study graphs in a quantum setting we first map them into quantum states known as graph states. To do so we follow the procedure described in the paper of Zhao \textit{et al.\ }~\cite{Zhao2016} which efficiently and uniquely encodes the graphs \cite{Hein2006,Mhalla2004,Hoyer2006,Cabello2011}, as set out here.  We then show it is possible to construct classes of measurements, derived from the Wigner function~\cite{PhysRevLett.117.180401,Rundle2016,Wigner1932}, which act as graph invariants.

\begin{figure}
\centering
\includegraphics[width=.95\linewidth,trim={0mm 0mm 0mm 0mm},clip]{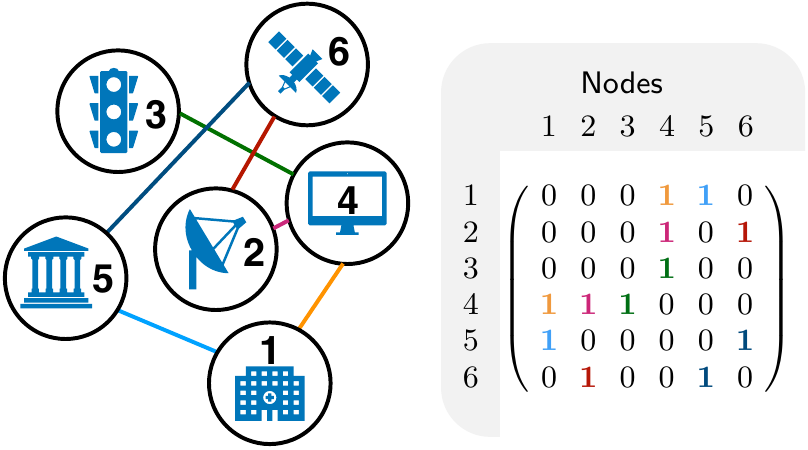}
\caption{Schematic showing an example network or graph and the associated adjacency matrix that represents it. The adjacency matrix is constructed by numbering each node and associating each row and column with a node. Ones are then entered into the positions corresponding to connected nodes and zeros otherwise. The set of adjacency matrices associated with isomorphic graphs are those that are formed from simultaneously permuting the matrices' rows and columns.}
\label{fig:graph}
\end{figure}

To obtain a graph state from a graph or network, first the adjacency matrix must be found as shown in \Fig{fig:graph}. 
A set of operators is then constructed from the adjacency matrix by replacing the elements of the adjacency matrix with Pauli matrices according to the following scheme: diagonal entries of the matrix are replaced by $\hat{\sigma}_x$; all other zeros are replaced by the identity operator, which we denote by $\hat{\sigma}_I$, and each `$1$' is replaced by a $\hat{\sigma}_z$. 
The tensor product is then taken between adjacent matrices within each row, resulting in a set of $N$ operators for an $N$-qubit system. 
Note that each qubit corresponds to a node in the graph.
For the adjacency matrix shown in \Fig{fig:graph}, the corresponding set of operators is given by;
\bel{eq:generators1}
\left\{\begin{array}{c}
\hat{\sigma}_x\otimes\hat{\sigma}_I\otimes\hat{\sigma}_I\otimes\textcolor[rgb]{.94,.6,.24}{\hat{\sigma}_z}\otimes\textcolor[rgb]{0.25,0.62890625,0.96875}{\hat{\sigma}_z}\otimes\hat{\sigma}_I\\
\hat{\sigma}_I\otimes\hat{\sigma}_x\otimes\hat{\sigma}_I\otimes\textcolor[rgb]{.8,.15,.47}{{\hat{\sigma}_z}}\otimes\hat{\sigma}_I\otimes\textcolor[rgb]{.715,.09,.03}{\hat{\sigma}_z}\\
\hat{\sigma}_I\otimes\hat{\sigma}_I\otimes\hat{\sigma}_x\otimes\textcolor[rgb]{.0,.435,.078}{\hat{\sigma}_z}\otimes\hat{\sigma}_I\otimes\hat{\sigma}_I\\
\textcolor[rgb]{.94,.6,.24}{\hat{\sigma}_z}\otimes\textcolor[rgb]{.8,.15,.47}{\hat{\sigma}_z}\otimes\textcolor[rgb]{.0,.435,.078}{\hat{\sigma}_z}\otimes\hat{\sigma}_x\otimes\hat{\sigma}_I\otimes\hat{\sigma}_I\\
\textcolor[rgb]{0.25,0.62890625,0.96875}{\hat{\sigma}_z}\otimes\hat{\sigma}_I\otimes\hat{\sigma}_I\otimes\hat{\sigma}_I\otimes\hat{\sigma}_x\otimes\textcolor[rgb]{.0,.29,.49}{\hat{\sigma}_z}\\
\hat{\sigma}_I\otimes\textcolor[rgb]{.715,.09,.03}{\hat{\sigma}_z}\otimes\hat{\sigma}_I\otimes\hat{\sigma}_I\otimes\textcolor[rgb]{.0,.29,.49}{\hat{\sigma}_z}\otimes\hat{\sigma}_x\\
\end{array}
\right\}
=
\left\{\begin{array}{c}
\hat{G}_1\\
\hat{G}_2\\
\hat{G}_3\\
\hat{G}_4\\
\hat{G}_5\\
\hat{G}_6\\
\end{array}
\right\}.
\ee
These operators, $\hat{G}_i, ~i\in \{1,\dots,N\}$, are known as group generators, since their products form a finite abelian group of $2^N$ operators $\hat{g}_k, ~k\in \{1,\dots,2^N\}$. 
Adding the group elements together and normalising gives the density matrix of the corresponding graph state. 
The density matrix can also be expressed in a factorised form using only the generators as shown on the right~\cite{Rocchetto2017}:
\bel{eq:DMinGens}
\hat{\rho}=\frac{1}{2^{N}}\sum_{k=1}^{2^N}\hat{g}_k=\frac{1}{2^{N}}\prod_{i=1}^N\left(\Bid+\hat{G}_i\right).
\ee

It is worth noting that graph states are a subset of stabilizer states~\cite{IandM}. They are pure states that are said to be fixed by the generators, i.e. $\hat G_i \hat \rho=\hat \rho$.
It is the form of the states which makes using graph states so promising for producing effective graph invariants. This is because all the information about the graph is in the state and every group element is accessible either individually or simultaneously via measurements. 
Hence all the information is available and it is just a question of generating schemes for efficiently accessing the relevant information.

Zhao~\cite{Zhao2016} also covers how to experimentally construct the states expressed in \Eq{eq:DMinGens}. 
First, each node is associated with a qubit prepared in the $+1$ eigenstate of $\hat{\sigma}_x$; $\ket{+}$. 
Controlled-Z gates are then applied between any two qubits whose corresponding nodes are connected in the graph. 
For a fully connected quantum computer this procedure requires $\mathcal{O}(N^2)$ operations. However, it has been shown by Zhao that it is theoretically possible to construct graph states with $\mathcal{O}(N)$ operations with the use of an oracle. 

We note that it is also possible to implement the procedure on hardware with limited connectivity provided a path of connections can be found between any two qubits connected in the graph. Regarding the current IBM machines, a CZ gate is not provided as required in the algorithm to generate graph states. This problem is easily overcome by adding appropriate Hadamard gates to the CNOT gates. In cases where there is less connectivity and there is no direct CNOT gate between two qubits, it is possible to `skip' a qubit with a sequence of CNOT gates, provided both of the qubits have a connection to a shared qubit. Although possible, the number of gates needed to `skip' higher numbers of qubits grows quickly, resulting in higher decoherence in the algorithm. A similar sequence of gates is also possible with CZ gates with a similar outcome. Thus it is possible to implement our algorithm on various architectures, as long as a minimum connectivity is met.

\section{Methods}

\subsection{Underpinning theory}

Having described how to encode a graph into a graph state, we now discuss how to measure its quantum graph invariants. 
An observable can only be a quantum graph invariant if its measurement results do not depend upon the order in which the qubits are labeled. 
Two ways of achieving such a measurement are to treat all qubits identically or to measure each qubit individually and then discard ordering information by sorting the individual measurement results according to some arbitrary norm, for example by magnitude.
The former case is order invariant for any given state since the measurement results contain no qubit index information and therefore the order in which the qubits are labelled must have no effect.

If we consider a general measurement on $N$ qubits
\be 
\hat{M}=\bigotimes_{j=1}^N \hat m_j
\ee
then the former case corresponds to taking $\hat m_j = \hat m$ for all $j$ giving, 
\be 
\label{form1}
\hat{M_0}=\hat{m}^{\otimes N}.
\ee
While, as an example of the second case, we consider the situation in which the observable being measured on one arbitrary qubit $\hat m_1$ is different to the observable being measured on all other qubits $\hat m_0$ giving, 
\be 
\label{form2}
\hat{M}_1
=\bigotimes_{j=1}^N
 \br{\delta_{jq} \hat{m}_1
+(1-\delta_{jq})\hat m_0}.
\ee

If this measurement is repeated for all $q\in\{1,...,N\}$, the expectation values lead to a quantum graph invariant once the values are sorted as shown here; 
\be
\EX{\boldsymbol{\hat{M}}}=\textrm{sort} \Bigg(\EX{\hat{M}^{(q=1)}_1},\EX{\hat{M}^{(q=2)}_1},\cdots, \EX{\hat{M}^{(q=N)}_1}\Bigg).
\ee
In either case the expectation values of these operators may be calculated in a similar manner given below.

Expanding \Eq{eq:DMinGens} the graph state may be written in terms of its generators as
\begin{equation}
\begin{split}
&\hat{\rho}=2^{-N}\prod_{i=1}^N\left(\Bid+\hat{G}_i\right)=2^{-N}\sum_{\underline{a}} \hat{G}_1^{a_1}\hat{G}_2^{a_2} \cdots \hat{G}_N^{a_N}\\
&=\frac{\sum_{\underline{a}} \Big ( \hat G_{1(1)}^{a_1}\hat G_{2(1)}^{a_2}\cdots \hat G_{N(1)}^{a_N} \otimes \cdots \otimes \hat G_{1(N)}^{a_1}\hat G_{2(N)}^{a_2}\cdots \hat G_{N(N)}^{a_N} \Big )}{2^N}
\end{split}
\end{equation}
where the sum is over all binary words $\underline{a}=[a_1,\cdots,a_N]$ of length $N$. Note that the second subscript specifies the space in which that part of the operator is acting. For example in \Eq{eq:generators1}, $\hat G^{1}_{2(4)}=\hat{\sigma}_z$ whilst $\hat G^{0}_{i(j)}=\hat{\sigma}_I$ for all $i,j$. The expectation value for a measurement of the form in \Eq{form1} is then
\be
\label{EM1}
\begin{split}
\EX{\hat{M}_0} &= \Tr (\hat{\rho} \hat{M}_0 )\\
&=2^{-N}\sum_{\underline{a}}\bigotimes_{j=1}^N \Tr(\hat G_{1(j)}^{a_1}\hat G_{2(j)}^{a_2}\cdots \hat G_{N(j)}^{a_N}\hat{m}).
\end{split}
\ee
For graph states, the $\hat G_{i(j)}$ correspond to a mapping of the adjacency matrix $A$, according to $\hat G_{j(j)} = \hat{\sigma}_x, \hat G_{i(j)} = \hat{\sigma}_I$ if $A_{ij} = 0$ and $\hat G_{i(j)} = \hat{\sigma}_z$ if $A_{ij} = 1$. Consequently a term from the product in \Eq{EM1} contains either one factor of $\hat{\sigma}_x$ (if $a_j=1$) or none (if $a_j$ = 0) together with a number of factors of $\hat{\sigma}_z$ equal to the edge count $v_j$ of node $j$; it may be written in the canonical form $\hat{\sigma}_x\hat{\sigma}_z^{v_j}$ by swapping $\hat{\sigma}_x$ with those $\hat{\sigma}_z$ terms on its left. In doing so, each swap introduces a change of sign as $\hat{\sigma}_x\hat{\sigma}_z=-\hat{\sigma}_z\hat{\sigma}_x$ and the final results may be reduced depending on whether the edge count $v_j$ is even (when $\hat{\sigma}_x\hat{\sigma}_z^{v_j} = \hat{\sigma}_x$) or odd (when $\hat{\sigma}_x\hat{\sigma}_z^{v_j}= \hat{\sigma}_x\hat{\sigma}_z = -\ui\hat{\sigma}_y$). As a result each trace in \Eq{EM1} evaluates to 

\be 
\label{term}
\begin{split}
&\Tr(\hat G_{1j}^{a_1}\cdots \hat G_{nj}^{a_n}\hat{m})\\
&=(-1)^{\frac{1}{2}(a_j (A\underline{a}^T)_j+a_jr_j)}\left \{ 
\begin{array}{cc}
\Tr (\hat{m} ) &(a_j,r_j) = (0,0)\\
\Tr (\hat{\sigma}_x\hat{m} ) &(a_j,r_j) = (1,0)\\
\Tr (\hat{\sigma}_y\hat{m} )&(a_j,r_j) = (1,1)\\
\Tr (\hat{\sigma}_z\hat{m} ) &(a_j,r_j) = (0,1)
\end{array} \right.\\
\end{split}
\end{equation}
Here the prefactor in \Eq{term} keeps track of the number of swaps and $r_j$ is the parity of the edge count $v_j$, \emph{i.e.} $r_j = 1$ if $(A\underline{a}^T)_j$ is odd and is 0 otherwise.

What then remains is to choose measurements $\hat m,\hat m_0$ and $\hat m_1$ as to provide a means of distinguishing non-isomorphic graphs. We consider a possibility for each case in the following sections. In  section~\ref{SECEA} measurements from the equal-angle slice of the Wigner function are used which have the form in \Eq{form1} and treat no qubits differently. 
This results in a natural graph invariant which identifies more than $99.8\%$ of the graphs we have tested,  outperforming all of the classical invariants we consider. 
In  section~\ref{SECAN} we consider a measurement distinguishing one qubit as in \Eq{form2}. 
The resulting invariant can identify all graphs we have tested when combined with the eigenvalues of the adjacency matrix. 
Finally, in section~\ref{SECDIAN},  we extend this scheme to the measurement of two qubits with an operator other than $\hat m_0$, giving an invariant which has, on its own, distinguished all graphs we have tested.

\subsection{Measurements which distinguish no qubits: the equal-angle slice of the Wigner function}
\label{SECEA}
\begin{figure*}[!tb]
\centering
\includegraphics[width=0.95\textwidth]{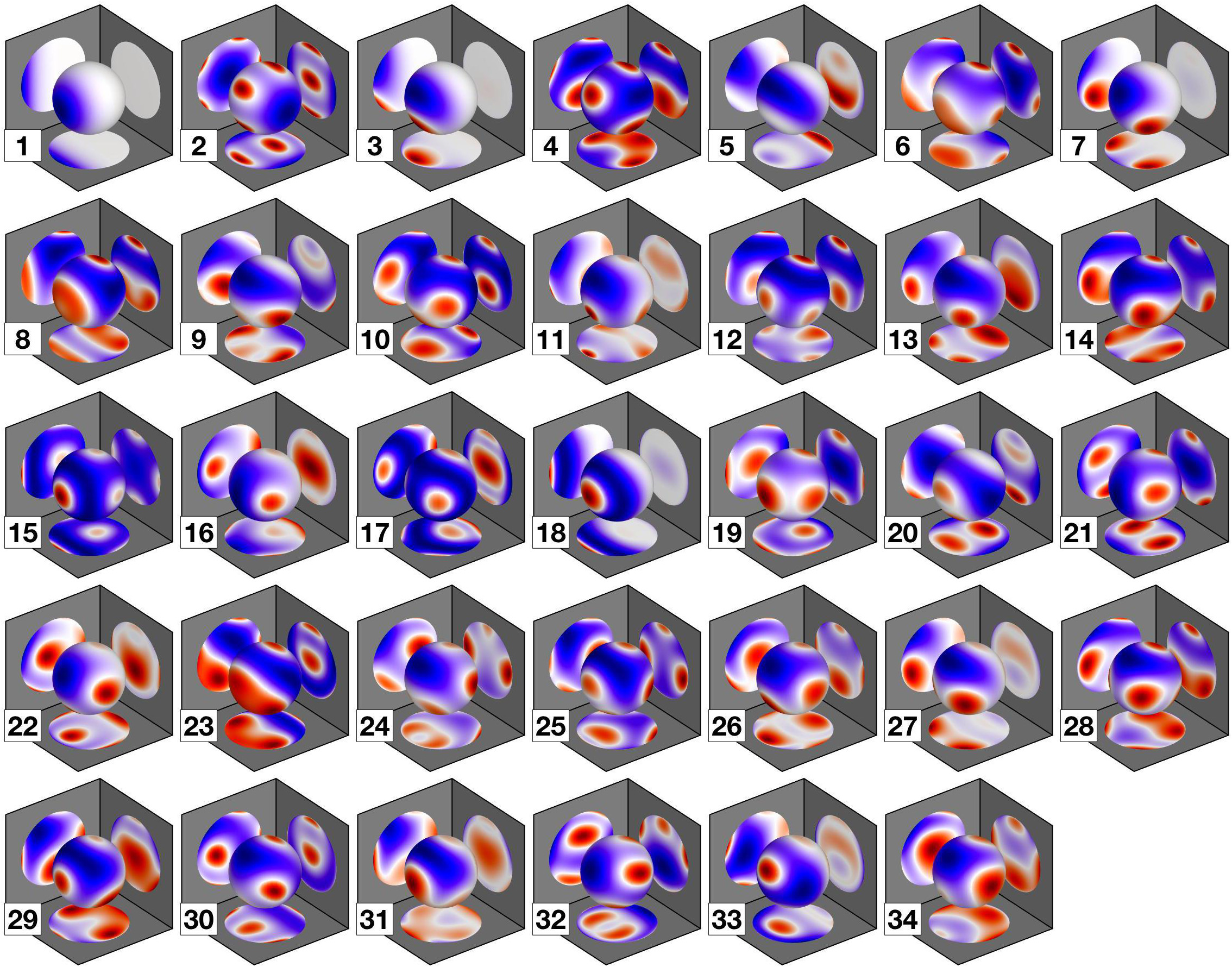}
\caption{We show the Wigner functions for the set of all five-node graphs that are not 
isomorphic. Note that the colour maps have been scaled for each figure to maximise the feature clarity. We have done this to better enable the reader to see the functional form of each graph (this does however mean that a direct comparison between plots in terms of magnitude is not possible). We have also computed the Wigner functions for all graphs of fewer than ten nodes and can use them to identify all graphs of fewer than eight nodes. Graphs with eight or nine nodes can still be efficiently identified by using anagraph measurements which have the form given in \Eq{eq:ColM}.
\label{fig:Wigners}
}
\end{figure*}

\begin{figure}
\centering
 \includegraphics[width=0.95\linewidth,trim = {0cm, 0cm, 0cm, 0cm},clip = true]{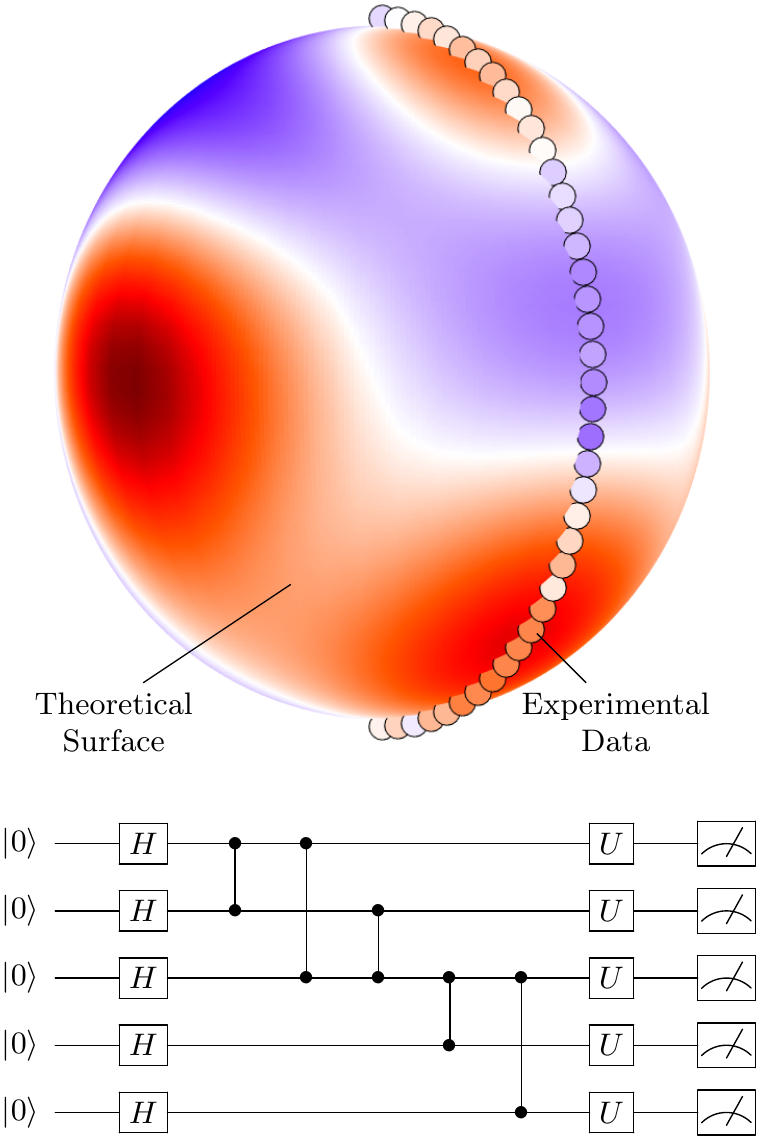}
\caption{\label{fig:IBM}Figure showing the equal-angle Wigner function (given by $\EX{\hat\Pi_{\frac{1}{2}}\br{\boldsymbol\theta,\boldsymbol\phi}}$) and construction circuit for graph number $34$ in~\Fig{fig:Wigners}. The sphere is the theoretical plot and the discs on its surface show the experimental data, measured at that point on the equal-angle sphere using IBM's 5 qubit machine via quantum experience. The unmarked gates represent Controlled-Z gates. We note that the IBM experience machines are still in development and that the deviations between theory and experiment seen here are to be expected. However, we were able to use the simulator in the IBM experience to obtain exact agreement with our own theoretical predictions, showing the process works in principle.}
\end{figure}

To perform identical measurements on all qubits we used the equal-angle slice of the Wigner function which has previously been used for state characterization~\cite{Rundle2016}. 
This lead us to speculate that it could be used to identify graph states. 
The results in~\Fig{fig:Wigners} where we show the equal-angle slice of the Wigner function for all 34 non-isomorphic 5-node graphs supported this speculation~\footnote{We note for completeness that we have verified for a selection of graphs that their isomorphisms do indeed produce \IsoWigner plots.}.
We show in~\Fig{fig:IBM} example experimental data calculated on the IBM quantum experience. 
We followed the procedure in \cite{Rundle2016} to directly measure points on the equal-angle Wigner function using IBM’s Quantum Information Software Kit (QISKit) that \emph{``is a software development kit (SDK) for working with OpenQASM and the IBM Q experience (QX)''} ~\cite{ibm}.

\begin{figure}
\includegraphics[width=0.95\linewidth,trim = {0cm, 0cm, 0cm, 0cm},clip = true]{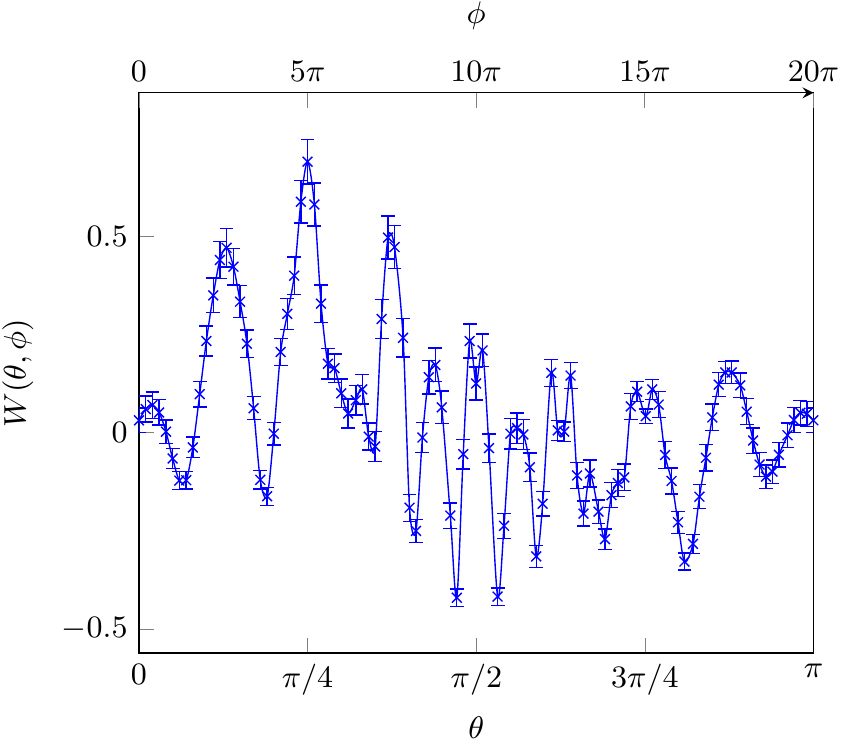}
\caption{\label{fig:EASprial}Figure showing 1 standard deviation about the expectation value of the equal-angle Wigner function for graph number $34$ in~\Fig{fig:Wigners}. The data points are distributed over a spiral covering the sphere. The standard deviation is theoretically calculated based upon 1000 preparations of the state. This distribution of points was used to facilitate a plot, but not for comparing states since it is non-uniform.}
\end{figure}

A detailed discussion of spin Wigner functions goes beyond the scope of this paper - please see Refs~\cite{PhysRevLett.117.180401,Rundle2016,Rundle2018} for a full discussion. In brief, a spin Wigner function for a set of $N$ qubits may be given in the form
\bel{TraceSpinDispParity}
W(\boldsymbol{\theta},\boldsymbol{\phi}) = \EX{\hat{\Pi} (\boldsymbol{\theta},\boldsymbol{\phi})} =\Trace{ \hat{\Pi}(\boldsymbol{\theta},\boldsymbol{\phi})\, \hat{\rho} }.
\ee 
Here $\boldsymbol{\theta}=(\theta_1,\dots,\theta_N)$ and $\boldsymbol{\phi}=(\phi_1,\dots,\phi_N)$ are the sets of coordinates of the Bloch sphere, associated with each qubit, and $\hat{\Pi}(\boldsymbol{\theta},\boldsymbol{\phi})$ is a rotated generalised parity operator given by
\bel{eq:Pi}
\hat{\Pi}(\boldsymbol{\theta},\boldsymbol{\phi})=\bigotimes_{i=1}^N   \hat{\Pi}_{\Half}^{(i)}(\theta_i,\phi_i).
\ee 
All the $\hat{\Pi}_{\frac{1}{2}}^{(i)}$ have the same form and are given by~\cite{Rundle2016}

\begin{equation}
\hat{\Pi}_{\frac{1}{2}}^{(i)}= \frac{1}{2}\left (\begin{array}{cc}
1+\sqrt{3}\cos\theta_{i}&-\sqrt{3}\sin\theta_{i} \exp (\ui \phi_{i})\\
-\sqrt{3}\sin\theta_{i} \exp (-\ui \phi_{i})&1-\sqrt{3}\cos\theta_{i}
\end{array} \right ).
\end{equation}
The equal-angle slice, required for treating all qubits equivalently, is then obtained by setting $\theta_i=\theta$ and $\phi_i = \phi$ for all $i$. This gives an operator in a form equivalent to that in \Eq{form1}.
Inserting $\hat\Pi(\theta,\phi)$ into \Eq{EM1} gives the equal-angle slice of the Wigner Function as
\be 
\begin{split}
W(\theta,\phi)&=\sum_{\underline{a}}(-1)^{\frac{1}{2}(\underline{a}A\underline{a}^T+\underline{a}\cdot\underline{r})} \Tr (\hat{\sigma}_y\hat{\Pi}_{\frac{1}{2}})^{\underline{a}\cdot \underline{r}}\times \\
&\Tr (\hat{\sigma}_x\hat{\Pi}_{\frac{1}{2}})^{(\underline{1}-\underline{r})\cdot \underline{a}}\Tr (\hat{\sigma}_z\hat{\Pi}_{\frac{1}{2}})^{(\underline{1}-\underline{a})\cdot \underline{r}}\\ 
\end{split}.
\ee
Evaluating the trace terms as
\be
\begin{split}
\Tr(\hat{\Pi}_{\frac{1}{2}})&=1\\
\Tr(\hat{\sigma_x}\hat{\Pi}_{\frac{1}{2}})&=-\sqrt{3}\sin\theta \cos \phi\\
\Tr(\hat{\sigma_y}\hat{\Pi}_{\frac{1}{2}})&=\sqrt{3}\sin\theta \sin \phi\\
\Tr(\hat{\sigma_z}\hat{\Pi}_{\frac{1}{2}})&= \sqrt{3}\cos\theta
\end{split}
\ee
leads to the final expression
\begin{equation}
\label{VinceEAWF}
\begin{split}
W(\theta,\phi)&=\sum_{\underline{a}}(-1)^{\frac{1}{2}(\underline{a}A\underline{a}^T-\underline{a}\cdot\underline{r})+\underline{1}\cdot \underline{a}} \sqrt{3}^{\underline{1}\cdot(\underline{a}+\underline{r})+\underline{a}\cdot \underline{r}} \\
&\times \cos^{(\underline{1}-\underline{a})\cdot \underline{r}}\theta \sin^{\underline{1}\cdot\underline{a}}\theta \cos^{(\underline{1}-\underline{r}) \cdot \underline{a}}\phi \sin ^{\underline{a}\cdot \underline{r}}\phi\\
&=\sum_{\underline{a}} C_{\underline{a},\underline{r}} \, x^{(\underline{1}-\underline{r}) \cdot \underline{a}} \, y^{\underline{a}\cdot \underline{r}} \, z^{(\underline{1}-\underline{a})\cdot \underline{r}}
\end{split}
\end{equation}
where the second expression comes from using the notation $x=\sin\theta\cos\phi$, $y=\sin\theta\sin\phi$, $z=\cos\theta$ and
\be 
C_{\underline{a},\underline{r}}=(-1)^{\frac{1}{2}(\underline{a}A\underline{a}^T-\underline{a}\cdot\underline{r})+\underline{1}\cdot \underline{a}} \sqrt{3}^{\underline{1}\cdot(\underline{a}+\underline{r})-\underline{a}\cdot \underline{r}}.
\ee
 
It is interesting to note that by analyzing coefficients of the terms in the polynomial in \Eq{VinceEAWF}, it is possible to determine a number of permutation invariant properties of the adjacency matrix, such as the degree sequence. This supports the possibility that equal-angle Wigner functions could fully encode the adjacency matrix (up to permutations), and in turn be used to distinguish all graph states.

However, as shown in \Tab{Tab1} we have found sets of non-isomorphic graphs which have the same equal-angle slice of the Wigner function. 
We call graphs with the same equal-angle slice \emph{equiumbral}~\footnote{meaning of equal shadow}. Thus isomorphic implies equiumbral but the converse does not hold. 
Despite this it is worth noting that the equal-angle slice of the Wigner function is dependent upon enough information in the states to identify more than $99.8\%$ of the graphs we have tested, significantly outperforming all the classical methods we have considered as shown in ~\Tab{Tab1}.

More generally we have found that any observable of the form in \Eq{form1} will not be able distinguish all graph states, despite their use in permutation invariant tomography \cite{PhysRevLett.105.250403}.

This is due to the existence of non-isomorphic graphs which share the same number of each type of Pauli operator within each group element. Since these graph states are not isomorphic they cannot be made equal by permutations of the whole group, which is equivalent to permuting nodes. However, they can be made identical by applying permutations to the operators in a subset of the group elements. Thus we find that `equal-angle' or `permutation invariant' measurements, are actually invariant for a larger class of possible `reorderings' including the partial permutations described above, and do not form an exclusively permutation invariant measurement (this suggests it would be more accurate to refer to permutation invariant tomography as order invariant instead).

This problem is overcome in the following sections with measurements which are sensitive to partial permutations of the group.

We now consider the efficiency with which this method can be used to check if graphs are isomorphic.
 If ensemble measurements at $P$ points are required to distinguish two \noniso graphs, then using the $\mathcal{O}(N)$ construction of the state, the whole procedure will require $\mathcal{O}(PN)$ operations. Determining with certainty that two graphs are equiumbral will require at most $P=(N^2+3N+2)/2$ ensemble measurements per graph. (In our work the measurements were distributed evenly over the Bloch sphere. This has been shown to be close to optimal~\cite{PhysRevLett.105.250403}.) 
Since non-isomorphic graphs may be distinguished in far fewer than $N^2$ ensemble measurements, the total procedure will require between $\mathcal{O}(N)$ and $\mathcal{O}(N^4)$ gate operations and ensemble measurements, dependent upon the graphs to be compared and the method of constructing the states. However, this analysis assumes negligible error in the ensemble measurements, which is realistic for small systems as shown in \Fig{fig:EASprial}. Assuming a measurement process similar to that used in IBM's quantum computers,  the number of preparations required per ensemble to distinguish pairs of graphs based on a single point appears to grow exponentially. This is due to the functions becoming similar as graph size increases. As such, an equal-angle approach may require alternative measurement procedures to effectively obtain the functions with sufficient clarity. Whether alternative procedures are possible which are significantly more efficient at distinguishing these functions is left to further work.

\begin{table*}
\caption{\label{Tab1} Rows 1, 5, 6 and 7 show values derived from ``The On-Line Encyclopedia of Integer Sequence"\cite{OEIS} giving, for each graph up to 9 nodes the number of; graphs (A000088), the completeness gap of degree sequences, i.e. the number of graphs minus the number of degree sequences (A004251), the number of graphs minus the number of graphs with a unique Tutte polynomial (A243049), and the completeness gap of eigenspectra (A099883). In each row we show the `completeness gap' which is determined as the number of graphs minus the number of distinct outcomes for a given invariant. Thus it gives an idea of how far an invariant is from being complete, i.e. able to distinguish all graphs. Rows 2, 3 and 4 show the completeness gaps using our quantum methods alone. Notably, in row 2 we have found we can distinguish at least all graphs with less than 10 nodes by performing anagraph measurements on 2 qubits simultaneously. Finally rows 8, 9 and 10 show the completeness gaps when combining methods. As shown in row 9 we find that combining anagraph measurements with eigenspectra allows for all graphs of less than 10 nodes to be distinguished. 
}
\begin{tabular}{|c|lccccccccc|}
\hline
Row \#&Number of Nodes&1&2&3&4&5&6&7& 8& 9\\
1&Number of graphs&
1&2&4&11&34&156&1,044& 12,346& 274,668\\
2&Completeness gap of dianagraph values &0&0&0&0&0&0&0&0&0\\
3&Completeness gap of equal-angle Wigner functions&0& 0& 0& 0& 0& 0&0& 14& 222\\
4&Completeness gap of anagraph values $(\alpha =1)$&0& 0& 0& 0& 0& 0& 0& 18& 1,174 \\
5&Completeness gap of eigenspectra&0& 0& 0& 0& 1& 4& 38& 661& 242,620 \\
6&Completeness gap of Tutte Polynomials&0& 0& 0& 4& 15& 84& 548& 5,629& 90,776\\
7&Completeness gap of degree sequences &0 &0 & 0 & 0 & 3 & 54 & 702 & 11,133 & 270,307 \\
\hline
8&Completeness gap of eigenspectra \& equal-angle Wigner functions&0& 0& 0& 0& 0& 0&0& 0& 18 \\
9&Completeness gap of eigenspectra \& anagraphs values $(\alpha =1)$&0& 0& 0& 0& 0& 0&0& 0& 0 \\
10& Completeness gap of $(\alpha =1)$ anagraphs \& equal-angle Wigner functions&0& 0& 0& 0& 0& 0&0& 2& 3 \\
\hline
\end{tabular}
\end{table*}

\subsection{Measurements which distinguish a single qubit}
\label{SECAN}
Having established a class of measurements which depend upon the group elements in their entirety it is natural to try to determine information from the group elements which corresponds mainly to individual qubits.  This can be done with an observable in the form of the second case, as given in \Eq{form2}, by taking

\bel{m0} 
\hat{m}_0=\hat\sigma_I+\alpha(\hat\sigma_x+\hat\sigma_y+\hat\sigma_z)
\ee
(where $\alpha$ is an adjustable parameter used to reduce the variance in measured results) and taking $\hat m_1$ in turn to be $\hat\sigma_I$, $\hat\sigma_x$, $\hat\sigma_y$ and $\hat\sigma_z$, giving

\bel{eq:ColM}
\hat{M}_i^k
=\bigotimes_{j=1}^N
 \left(\delta_{jk} \hat{\sigma}_i
+(1-\delta_{jk})
\hat m_0\right)
\ee
for qubit $k$ and Pauli matrix $i$.

Substituting $\hat m_0$ for $\hat m$ into \Eq{term}, the trace terms for $j\neq k$ are respectively $[2,2\alpha,2\alpha,2\alpha]$, while the terms 
\be 
[ \Tr(\hat{m}_1), \Tr(\hat{\sigma_x}\hat{m}_1), \Tr(\hat{\sigma_y}\hat{m}_1), \Tr(\hat{\sigma_z}\hat{m}_1)]
\ee
will give $[2,0,0,0], [0,2,0,0], [0,0,2,0]$ and $[0,0,0,2]$, respectively for $\hat{m}_1=\hat{\sigma}_I,\hat\sigma_x,\hat\sigma_y$ and $\hat\sigma_z$. As a consequence, the factor of $2^N$ cancels the normalization of the graph state and measuring $\hat{m}_1 = \hat{\sigma}_I,\hat\sigma_x,\hat{\sigma}_y$ and $\hat{\sigma}_z$ on qubit $k$, and $\hat{m_0}$ on the rest will yield a matrix $\mathcal{M}$ whose 4$N$ values are dependent upon the number of occurrences of the operator $\hat{\sigma}_i$ in the $k^\mathrm{th}$ space of the group elements, given by;

\be 
\begin{split}
\left [ \begin{array}{c}
M_I^{k}\\
M_x^{k}\\
M_y^{k}\\
M_z^{k}
\end{array} 
\right ]=&\sum_{\underline{a}}(-1)^{\frac{1}{2}(\underline{a}A\underline{a}^T+\underline{a}\cdot\underline{r})}\times\\
&\alpha^{\underline 1\cdot(\underline a + \underline r)-\underline a \cdot \underline r}
\left [\begin{array}{c}
(1-a_k)(1-r_k)\\
a_k(1-r_k)\\
a_kr_k\\
(1-a_k)r_k
\end{array}
\right ] 
\end{split}.
\ee
We fix the parameter $\alpha=1$ for now, returning to its consideration again later. The resulting $4N$  elements of $\mathcal{M}$ are then integers and, when sorted by column $k$, provide a matrix which is invariant under any permutation of the qubits. 


As an example of how to perform such a measurement experimentally, consider `counting' the number of $\hat \sigma_x$ operators on the second qubit. Thus our operator will have a $\hat \sigma_x$ in the space of the second qubit. In the space of all other qubits the operator $\hat\sigma_I+\hat\sigma_x+\hat\sigma_y+\hat\sigma_z$ is measured. This corresponds to measuring $\sqrt{3}\hat U(\theta,\phi)\hat\sigma_{Z}\hat U^{\dagger}(\theta,\phi)$, with $\theta = -\arctan(\sqrt{2})$, $\phi = -\pi /4$ and $\hat U(\theta,\phi)=\exp(\ui\phi\Sz/2)\exp(\ui\theta\hat\sigma_y/2)$ 
\footnote{We note that  it is possible to adjust the above arguments to efficiently obtain from the graph-state density matrix the generators and therefore the adjacency matrix from a given graph state. This will be covered in a future paper, see also~\cite{Montanaro2017} for identifying stabilizer states.}.

Unfortunately the sorted matrix $\mathcal{M}$ with elements $M_i^k$ can be the same for \noniso graphs. Two graphs with identical matrices $\mathcal{M}$ we term \emph{anagraphs}, since they appear to contain the same number of each Pauli operator in the space of each qubit. We therefore have that graphs which are isomorphic will be anagraphs but not the converse.

Despite this we have found that graphs which are anagraphs and isospectral are isomorphic in all cases we have considered. It is not known whether this holds generally. However, in the next section we introduce a measurement which is robust against the failure modes of both the equal angle and single qubit approaches.

We now consider the efficiency with which anagraph measurements can be used to check if graphs are isomorphic. Since each operator of the four operators $\hat\sigma_{I},\hat\sigma_x,\hat\sigma_y$ and $\Sz$ must be measured on each qubit there are $4N$ ensemble measurements to perform.
Using the $\mathcal{O}(N)$ construction of the state, the whole procedure will require  $\mathcal{O}(4N^2)$ gate operations and ensemble measurements, assuming a constant number of preparations. If this assumption holds as system size increases, anagraphs will provide a powerful and efficient invariant. To check this assumption we now evaluate the theoretical signal to noise of anagraph operators.

\begin{figure}
\centering
\includegraphics[width=0.95\linewidth,trim = {0cm, 0cm, 0cm, 0cm},clip = true]{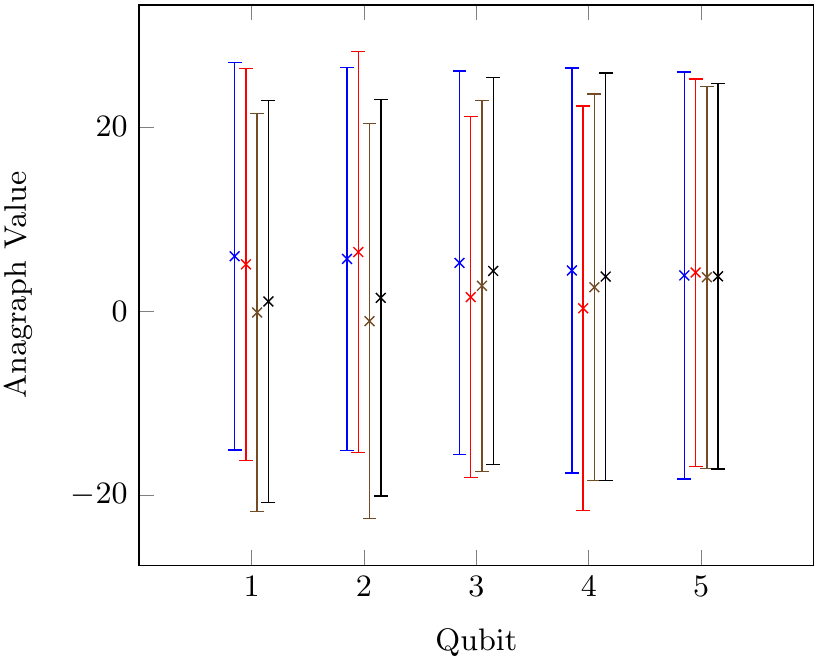}
\caption{\label{fig:ExpAna}Experimentally obtained anagraph values for the 5 node graph numbered $34$ in~\Fig{fig:Wigners}. The error bars show one standard deviation based upon $4096$ preparations. They can be reduced by varying the parameter $\alpha$ in \Eq{m0}. For each qubit the data points show the values for $\hat \sigma_{I}$, $\hat\Sx$, $\hat\Sy$ and $\Sz$ from left to right. The experimentally obtained variances closely match the theoretically obtained values, giving poor signal to noise as the number of qubits increases. Varying $\alpha$ improves signal to noise at the cost of making anagraph values of \noniso graphs more similar.}
\end{figure}

A value of $\alpha = 1$ has the advantage of providing a $4 \times N$ graph measure $\mathcal{M}$ consisting entirely of integer values, and consequently capable of delivering a robust means for the comparison of graphs states. Unfortunately, with this measure, the variance in measurement results grows exponentially with node number $N$, so that an exponential number of measurements is needed to obtain meaningful results. See \Fig{fig:ExpAna}. Nonetheless, it seems reasonable to suppose that other measures exists, dependent on $\mathcal{M}$, which possess better measurement statistics without losing the ability to distinguish graph states. One trivial example, which suggests such statistics exits, may be seen by altering the scale parameter $\alpha$ in the measurement operator $m_1$. For small $\alpha$ the signal to noise (SNR) can be shown to behave as roughly
\be 
\mathrm{SNR}=\frac{\langle M_{\hat{\sigma}}^k\rangle^2}{\Big \langle \Big(M_{\hat{\sigma}}^k-\langle M_{\hat{\sigma}}^k\rangle\Big)^2\Big \rangle} \sim \br{\exp(N\alpha^2)-1}^{-2}
\ee 
for Pauli operator $\hat{\sigma}$ and node $k$. Thus for a given value of node number $N$ one can significantly increase the SNR by choosing a value of $\alpha \ll N^{-1/2}$. We have shown in calculations that $\alpha \neq 1$ retains the ability to distinguish graph states, yet provides a much improved SNR.

\begin{figure}
    \centering
    \includegraphics[width=0.95\linewidth,trim = {0cm, 0cm, 0cm, 0cm},clip = true]{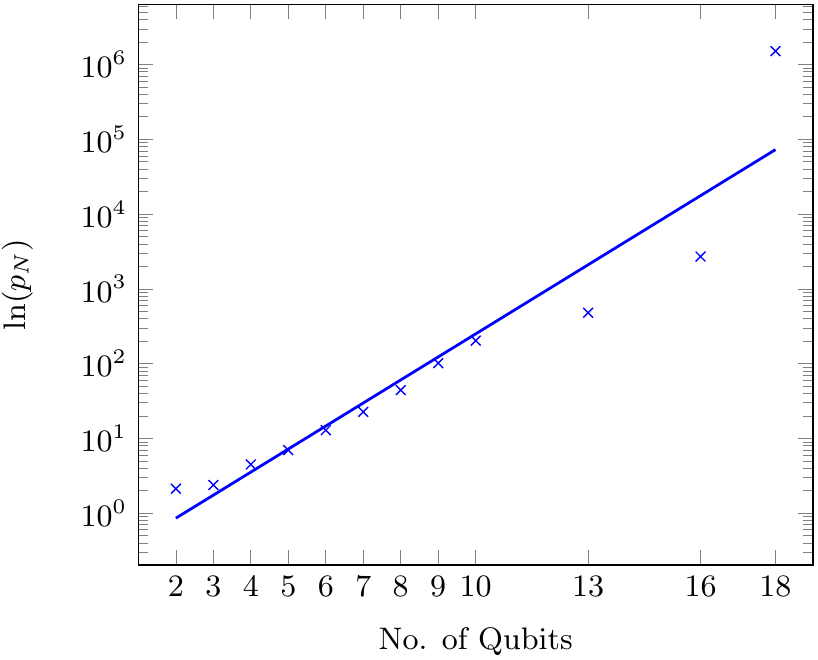}
    \caption{\label{fig:AnanPreps} Expected number of preparations $p_N$, to distinguish graphs using anagraph values. The value is the expected number of samples to have one standard deviation between results, based on up to 200 pairs of graphs per vertex count. The 13, 16 and 18 node values are based off only two graphs each, giving only a rough estimate of the scaling. }
\end{figure}

The cost of this reduction in signal to noise is that the anagraph values of \noniso graphs become more similar and therefore harder to distinguish. \Fig{fig:AnanPreps} shows the expected number of preparations to be able to distinguish graphs as a function of graph size. Whilst such a simple modification is insufficient to help distinguish graphs it remains a possibility that some modification of the measurements or data processing may be able to efficiently distinguish \noniso graphs. 

We have begun investigating operators which target only those group elements which are a product of a set number of generators. It appears these operators may require less preparations. Such operators will be the subject of future work. 
\subsection{Measurements distinguishing two or more qubits}
\label{SECDIAN}
A whole family of operators can be formed by varying the number of qubits which are measured with a different operator to $\hat m_0$. The equal-angle slice is an example with no qubits being measured with a different operator (although we rotate $\hat m_0$ to obtain more information about the state).
We then vary a single operator on individual qubits to give the anagraph values. 
Following a similar regime, but distinguishing two qubits at a time by measuring them with Pauli operators instead of $\hat m_0$, we find that at least all graphs with less than 10 nodes can be distinguished as well as all strongly regular graphs shown in row 2 of \Tab{Tab1} and \Tab{tab:SRGraphs}. This gives a measurement operator of the form
\bel{form3}
\hat M_2 = \bigotimes_{j=1}^N\br{\delta_{jq}\hat m_1 + \delta_{jp}\hat m_2 + \br{1-\delta_{jq}}\br{1-\delta_{jp}}\hat m_0},
\ee
provided $(q,\hat m_1)\neq(p,\hat m_2)$. $\hat m_2$ is chosen similarly to $\hat m_1$ as any of the Pauli matrices and $\hat m_0$ is as in \Eq{m0}. We call operators of this form dianagraph operators, and refer to the general family as polyanagraph operators.
To remain permutation invariant all possible pairs must be measured and in general for $k$ distinguished qubits there will be `$N$ choose $k$' sets which must be measured. Furthermore, if all possible combinations of Pauli matrices (including $\hat\sigma_{I}$) are measured on each qubit in a chosen set of $k$ qubits, the number of measurements required per set increases exponentially as $4^k$.

Whether this family of operators is applicable to solving the graph isomorphism problem is unclear, however they show promise in that the forms with $k=1$ and 2 have successfully distinguished all graphs we have been able to test. Based on this we conjecture that any pair of graphs with $N$ nodes will be distinguishable using measurements from the $N+1$ members of this family.

\begin{table}
\caption{\label{tabDegen}The table below shows the number of $9$-node graphs which share the same equal-angle Wigner function, anagraph values or eigenvalues. Note that all observed equiumbral graphs fail only in pairs, whilst anagraphs can share common values with more than one other graph. All $8$-node graphs which fail, only fail pairwise for all quantum methods.}
\begin{tabular}{|l|ccccccccc|}
\hline
Degeneracy(set size)&2&3&4&5&6&7&8&9&10\\
\hline
Equiumbral sets&222&0&0&0&0&0&0&0&0\\
Anagraph sets&345&13&0&0&1&0&1&0&0\\
Isospectral sets &25762&2015&551&95&37&1&2&0&2\\
\hline
\end{tabular}
\end{table}

\begin{table}
\caption{\label{tab:SRGraphs}The table below shows the classes of strongly regular graphs which have been distinguished using our invariants. The graphs were obtained from Spence's~\cite{TedSpence} online catalogue. The parameters correspond to the number of nodes, the degree of the nodes, the number of common neighbours adjacent nodes have and finally the number of neighbours non-adjacent nodes have in common~\cite{Godsil2001}. By definition graphs which permit these parameters are strongly regular.}
\begin{tabular}{|c|}
\hline
SR Graph Parameters\\
\hline
16-6-2-2\\
25-12-5-6\\
26-10-3-4\\
28-12-6-4\\
29-14-6-7\\
\hline
\end{tabular}
\end{table}


\section{Discussion of Results}

Direct classical simulation of these methods scales exponentially. Despite this we have found it tractable to numerically evaluate all $288,266$ graphs of fewer than ten nodes as well as all pairs of strongly regular graphs of up to 29 nodes. In each case finding both the equal-angle slice of the Wigner function and evaluating the anagraph values. 

A summary of our results is given in \Tab{Tab1},
 where our methods may be compared against classical graph invariants such as the number of isospectral graphs and Tutte polynomials. To determine the number of unique measurement results a hashmap was formed for each invariant. This involves mapping the results for each graph into a value known as a key. If two graphs have the same result they will have the same key but not otherwise. Associated with each key is the number of any graphs which have that key, where the graph's number is determined by Maple 2017.3's non-isomorphic graph generator. This method allows graphs with identical measurement results to be found quickly, as they will share the same key, whilst also allowing graphs with similar results to be found and compared by sorting the hashmap by the keys and then comparing adjacent entries.

The number of graphs which share any particular key is smaller for the quantum invariants than the classical invariants. The degeneracy of these measures for nine node graphs is shown in \Tab{tabDegen}.
For the special case of non-isomorphic strongly regular graphs which scale badly for classical algorithms such as Nauty, we show the classes we have successfully distinguished in \Tab{tab:SRGraphs}. We note that our invariants were also able to distinguish the pair of 24 node Mermin magic square graphs which were indistinguishable to the quantum approach of Asterias~\cite{Atserias2016}.

\subsection{Extensions to weighted graphs}

\begin{figure}
    \includegraphics[width=0.95\linewidth,
    trim = {2cm, 8cm, 2cm, 3cm},
    clip = true]
    {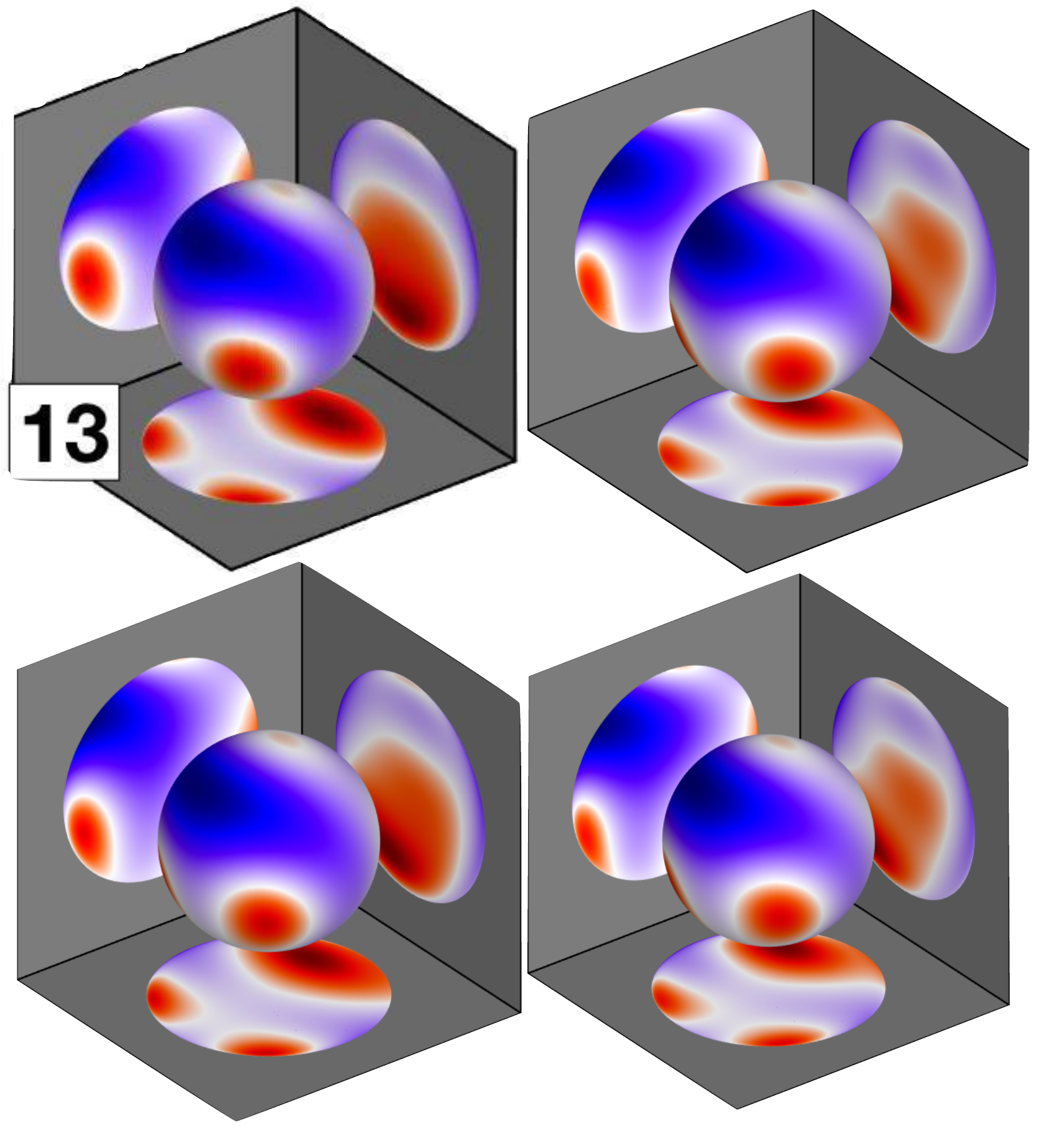}
\caption{\label{fig:Weighted}Shown above in the first row are the equal-angle Wigner functions of the $5$-node graph numbered $13$ in \Fig{fig:Wigners} and a weighted graph which is structurally similar to it ($\omega = 0.9\pi$ on all edges). In the second row we show randomly weighted graphs also with a similar form with average $\omega$ values of $0.95\pi$ and $0.9\pi$ respectively. Due to weighted connections the resultant functions are slightly distorted from that originally obtained on the top left. However, the topography of the function appears to be robust to small changes in the edge weights.}
\end{figure}

It should also be possible to apply our invariants to weighted graphs states, formed by the application of controlled phase gates as opposed to CZ gates~\cite{Hein2006} as in section \ref{sec:Constructing}. Denoting the phase angle by $\omega$, a controlled phase gate is given by
\be
\begin{bmatrix}
1 & 0 & 0 & 0\\
0&1&0&0\\
0&0&1&0\\
0&0&0&\mathrm{e}^{\ui\omega}
\end{bmatrix}.
\ee
Thus taking $\omega=0$, the identity matrix is obtaining giving no connection between the nodes, whilst for $\omega = \pi$, the CZ operator is obtained giving a full edge. Weighted edges can be obtained by taking intermediate values of $\omega$.
Equal-angle functions are order invariant for all states and as shown in \Fig{fig:Weighted}, small variations in the edge weights appear to correspond to small variations in the equal-angle Wigner functions. We have also performed preliminary checks on anagraph measurements finding that graphs with a small phase difference have similar anagraph values. This is shown in \Tab{tab:AnaWeight} where the anagraph values of a simple and similar randomly weighted graph are given. This is not surprising since the expectation values will be a continuous function of the phase angle guaranteeing that similar graphs will have similar expectation values. However, a full investigation is left to future work.

\begin{table}
\caption{\label{tab:AnaWeight}This table shows the anagraph values of simple graph and a structurally similar randomly weighted graph with noise applied to the edge weights. The average value of $\omega$ is $0.9\pi$. Each pair of rows shows the corresponding anagraph values from each graph type. The operator measured for that pair is shown in brackets. The columns represent distinct qubits on which the operator was measured. The order is determined by sorting the anagraph values.}
\begin{tabular}{|l|ccccc|}
\hline
($\hat\sigma_I$) Simple Graph & 6 & 6 & 6 & 6 & 8\\
Random Weighted Graph & 6.60 & 6.52 & 6.53 & 6.68 & 8.71\\
\hline \hline
($\hat\sigma_x$) Simple Graph & 2 & 2 & 6 & 6 & 4\\
Random Weighted Graph & 2.63 & 2.52 & 6.53 & 6.68 & 4.02\\
\hline \hline
($\hat\sigma_y$) Simple Graph & 6 & 6 & 2 & 2 & 0\\
Random Weighted Graph & 6.77 & 6.87 & 2.75 & 2.94 & 0.67\\
\hline \hline
($\hat\sigma_z$) Simple Graph & 6 & 6 & 6 & 6 & 8\\
Random Weighted Graph &6.10 & 6.19 & 6.30 & 5.80 & 8.69\\
\hline
\end{tabular}
\end{table}

\section{Conclusion and remarks}

In this work we have introduced a family of new graph invariants which can be determined on a quantum simulator/computer. 
We have also shown that these invariants can distinguish a greater proportion of non-isomorphic graphs than classical invariants. 
Our strategy makes use of the fact that a quantum state uniquely representing a $N$ node graph can be efficiently constructed using $N^2$ simple gate operations applied to an initial spin-coherent state on $N$ qubits. This renders all the information about the graph into a state which may then be arbitrarily probed for relevant information. It is our belief that this will provide many opportunities for the discovery or creation of novel and powerful graph invariants, providing a promising step in
understanding the graph isomorphism problem.

We have tested the viability of such an approach by applying our method on the IBM Q experience, and ,in good agreement with theory, found that extracting information with sufficient accuracy requires an exponentially large number of preparations. This is due to a trade off between either variance diverging with system size or values of different graphs becoming too similar to distinguish even with relatively simple operators. Whether it is possible to circumvent these problems remains an open question.
We have found that being order invariant, the equal-angle slice and similar operators are fundamentally unable to distinguish all graphs. Polyanagraphs were developed to only be permutation invariant, and in the case of dianagraphs may achieve this goal.  

Although our algorithms and the Weisfeiler-Lehman algorithms both perform graph canonisation, we believe there are key differences which make our algorithms distinct from their approach. The ability to distinguish strongly regular graphs shows our algorithms are not equivalent to the 2 dimensional Weisfeiler-Lehman algorithm. Our procedure is not based on iterative refinement, instead we canonically construct a unique graph state (guaranteeing measurable differences) and measure invariant properties from it. 
Furthermore, given the exponential resources on a quantum computer which are exploited by the algorithms, we conjecture that we may be accessing properties not observed in classical Weisfeiler-Lehman algorithms. Distinguishing the 80 node Cai-F\"{u}rer-Immerman graphs~\cite{Furer1987} would prove our algorithms are unrelated to Weisfeiler-Lehman's. However, it is not possible to evaluate such large graphs with current technology.

\begin{acknowledgments}
We acknowledge use of the IBM Q experience for this work. 
The views expressed are those of the authors and do not reflect the official policy or position of IBM or the IBM Q experience team.
R.P.R. is funded by the EPSRC [grant number EP/N509516/1].
T.T. notes that this work was supported in part by JSPS KAKENHI (C) Grant Number JP17K05569. 
S.J.D. acknowledges support from the Australian Research Council Centre of Excellence in Engineered Quantum Systems EQUS (Project CE110001013). 
We would also like to thank J.~Stark, D.~Bacon, S.~Severini and G.~M.~Fratangelo for their help suggesting example graphs to test our invariants. MJE would like to thank W.~J.~Munro for feedback and advice. 
MJE would like to thank Roberto Desimone and the Innovate UK \emph{``quantum algorithms for optimised planning and scheduling''} project for providing current context on the importance of the graph isomorphism problem to industry.
Please send queries on computational representation, modelling and optimisation to VMD, all other enquiries to MJE.
\end{acknowledgments}

\bibliographystyle{apsrev4-1}
\bibliography{refs}

\end{document}